\documentclass[aps, prl, a4paper, showpacs, twocolumn, superscriptaddress,10pt]{revtex4-1}

\usepackage{textcomp,hyperref,enumerate, url, cancel}   
\usepackage[margin=0.8in]{geometry}
\usepackage{color} 
\usepackage{amsmath,amssymb,amsthm,bbm,bm,nicefrac}   				
\usepackage{graphicx,epsfig}                                                               		
\usepackage{multirow}
\usepackage{subfigure}
\usepackage{slashed,ulem}

\newcommand{\ssection}[1]{\emph{#1.---}}

\newcommand{\ket}[1]{\left|#1\right\rangle}							
\newcommand{\bra}[1]{\left\langle#1\right|}

\newcommand{\ketbra}[2]{\left|#1\rangle\langle#2\right|}

\newcommand{\braket}[2]{\left\langle #1\lvert#2\right\rangle}
\newcommand{\abs}[1]{\left\lvert #1\right\rvert}

\newcommand{\emptybraket}[1]{\left\langle #1\right\rangle}

\newcommand{\be}{\begin{equation}} 							
\newcommand{\ee}{\end{equation}}
\newcommand{\ba}{\begin{align}}
\newcommand{\ea}{\end{align}}
\newcommand{\bematrix}{\left(\begin{matrix}}
\newcommand{\ematrix}{\end{matrix}\right)}
\newcommand{\Si}{\mathrm{Si}} 
\theoremstyle{definition}
\newtheorem{definition}[]{Definition}

\theoremstyle{theorem}
\newtheorem{theorem}{Theorem}

\theoremstyle{lemma}

\theoremstyle{proposition}

\theoremstyle{corollary}

\theoremstyle{observation}

\theoremstyle{remark}

\newcommand{\tr}[1]{\mathrm{Tr}\left[#1\right]}

\newcommand{\DB}[1]{{\color{black}#1}}

\begin{document}
 \title{Quantifying Causal Influence in Quantum Mechanics}
\author{Llorenç Escolà Farràs}
\affiliation{Institut f\"ur Theoretische Physik, Universit\"at T\"ubingen, D-72076 T\"ubingen, Germany. }
 \author{Daniel Braun}
\affiliation{Institut f\"ur Theoretische Physik, Universit\"at T\"ubingen, D-72076 T\"ubingen, Germany. }

\begin{abstract}
  We extend Pearl's definition of causal influence to the quantum
  domain, where 
  two quantum systems $A$, $B$ with
  finite-dimensional Hilbert space are embedded in a common
  environment $C$ and  propagated with a joint unitary $U$. {For finite dimensional Hilbert space of $C$,} we find the
  necessary and sufficient condition on $U$ for a causal influence of
  $A$ on $B$ and vice versa.   
 We introduce an easily
  computable measure of the
  causal influence
  and use it to study the causal influence of different quantum gates, {its} mutuality, {
and    quantum superpositions of different causal orders. 
    }
  {For 
    two
    two-level atoms dipole-interacting with a thermal
    bath of electromagnetic waves, 
    the space-time dependence of causal
    influence almost perfectly reproduces the one of 
    reservoir-induced entanglement. 
  }

\end{abstract}

\maketitle

\ssection{Introduction} To infer causes from effects
constitutes a key 
task of science. Classically, causal influence (CI) is defined between random variables (RVs)
that take certain values \DB{in a set of possible outcomes of randomized experiments. 
In a causal model, the RVs sit on vertices of a graph, and the 
forward-in-time-only CI in the classical world is represented by an arrow in a
directed acyclic graph.}  
According to Pearl \cite{BookOfWhy} {(see p.276)}, a
RV $x$ has a CI on another RV
$y$, if $y$ ``listens'' to $x$, meaning that there is a functional
relationship of the form  $y=f(x)+z$, where $f$ is some function and $z$ another RV. Apart from direct CI, correlations between $x$
and $y$ might arise additionally due to common causes. 
These can be
eliminated in practice by ``do-interventions'', where $x$ is 
randomly set by the experimenter and one examines if $y$ reacts.  {The corresponding (do-)probabilities, obtained by randomized controlled experiments 
  or via the do-calculus, 
  are the basis of Pearl's definition of CI as well as a measure of the average causal effect \cite{CausalityPearl}. }
  
Recently, there has been large interest in generalizing causal
analysis to the quantum world 
\cite{Gachechiladze2020, Oreshkov2012, Leifer2013, Brukner2014, Fitzsimons2015, Ried2015, Chiribella2015, Chaves2015, Pienaar2015, barrett2019quantum, Brukner2015, Costa2016, Pienaar2017,Ringbauer2016, Oreshkov2016, Branciard2016, Allen2017, SnchezBurillo2018, Milburn2018, Kbler2018, Hu2018, Milz2018, Guo2020, Oreshkov2019, Zych2019, foo2021thermality, Utagi2021, Barrett2021, costa2020nogo, Zhang2020}.  
One of
the most exciting perspectives 
is to superpose different temporal orders, and hereby
create ``indefinite causal order''.  To that end, process matrices
were introduced with separate input and
output Hilbert spaces of each laboratory, and the possibility to
``wire'' CI in 
opposite directions (e.g.~from the
output of Alice to the input of Bob or vice versa)
\cite{Oreshkov2012}.  The ``quantum switch'' was invented \cite{Chiribella2012, Chiribella2013} and
experimentally verified 
\cite{Procopio2015, Rubino2017,Goswami2018,CastroRuiz2018,Wei2019}. 
Here a
control qubit 
enables opposite temporal order of two quantum gates, 
which can improve the communication capacity of quantum channels,
and lead to a computational or metrological advantage \cite{Chiribella2012,Ebler2018,mukhopadhyay2018superposition}. Common to most of these
developments is, however, that the actual CI remained
unexplored, and the ``indefinite causal order'' refers to indefinite
(i.e.~superposed, or mixed) temporal orders. In order to 
study superpositions of 
different causal orders $x\to y$ and $y\to x$, one needs 
to define what is meant with a
CI $x\to y$ in quantum mechanics (QM).
In \cite{Allen2017,barrett2019quantum,Barrett2021} definitions of CI in QM based on the Choi-Jamio{\l}kowski representation of a unitary channel propagating Alice's and Bob's system were given, following earlier work in \cite{schumacher_locality_2005,beckman_causal_2001,eggeling_semicausal_2002}. Here we give a clear operational definition of CI based simply on density matrices,
generalizing the one by Judea Pearl from classical
statistical analysis \cite{BookOfWhy,CausalityPearl}.  We prove a theorem that gives the
necessary and sufficient condition for CI in QM and
introduce an easily computable measure of CI.  We use it
to analyse the CI of standard quantum gates, examine
some of the measure's statistical properties, as well as a quantum
causal switch that superposes two different CIs.
At the example of a 2-spin-boson model, we examine propagation of causal
influence.  We find that substantial CI arrives only far
behind the light-cone, and, surprisingly, almost perfectly in sync with
reservoir-induced entanglement.

We take a conservative approach based on standard QM \DB{(using density matrices rather than process matrices)} and the admission
of 
do-interventions.
Probability distributions of classical RVs are replaced 
by quantum states, since observables have, in general, no determined value
in QM until they are measured \cite{BellTheorem,Hensen2015,Wiseman2015}.  On the other hand,
quantum states encode all that can be known
about a quantum system, and hence it is  natural to base a theory of
CI in QM on states: ``causal influence''
in QM will be understood  in the sense
that the final quantum state of the causally influenced system
``listens to'', i.e.~depends on the
initial quantum state of the influencer.  Below we make this idea
mathematically precise {and introduce a measure of CI that we then explore}. 
\DB{In principle one could base a definition of CI also on correlation functions. If all correlations are included, this is equivalent to using the quantum state, but at the same time appears to be more cumbersome and less fundamental: fundamentally the world is quantum, and quantum computers will one day
probably be able to exchange quantum information without doing measurements.  This
motivates our attempt to base a definition of CI in
quantum mechanics directly on quantum states.}

\ssection{No-causal-influence condition} Consider two quantum systems $A$
(Alice) and $B$ (Bob) described by their respective density matrices
$\rho^A$ and $\rho^B$ and a third system $C$, the joint environment,
with a fixed initial state $\rho^C$ that can 
e.g.~correspond to the physical
system that propagates the CI and creates an effective
interaction, but also leads to decoherence.  The joint quantum system
lies in the Hilbert space
$\mathcal{H}_A\otimes\mathcal{H}_B\otimes\mathcal{H}_C$, with
dim$(\mathcal{H}_J)=d_J$, $J\in\{A,B,C\}$, with finite $d_A$, $d_B$,
whereas $d_C$ can be infinite. For further purposes, we define
$[d_J]:=\{0,...,d_J-1\}$. 
\begin{definition}\label{DefCausalInfluence} 
  Let $A,B,C$ at initial time
  $t_0$ be in the state $\rho(t_0)=\rho^B\otimes \rho^{AC}$, where
  $\rho^B$ is a state set by Bob in a do-intervention.  We say that $B$ at time $t_0$ 
  does not 
  causally 
  influence $A$ at time $t$ for a given initial $\rho^A$ if and
only if the reduced state $\rho'^A$ of the system $A$ after the
propagation from $t_0$ to $t$ is independent of $\rho^B$ for any
density matrix $\rho^B$.
    If this condition is fulfilled for any initial $\rho^A$, we say
    that system $B$ at time $t_0$ does not causally influence system
    $A$ at time $t$, shortly denoted by $B(t_0)\nrightarrow
    A(t)$. Otherwise we say that $B$ at time $t_0$ causally influences
    $A$ at time $t$, denoted by $B(t_0)\rightarrow A(t)$. Analogously
    one defines $A(t_0)\nrightarrow B(t)$ and $A(t_0) \rightarrow
    B(t)$, {leading to a total of four possible cases}. 
  \end{definition}
Note that if  one wants to
find out whether there is a CI from $A$ at time $t_0$ to $B$ at time
$t$, one needs an initial state $\rho(t_0)=\rho^A\otimes
\rho^{BC}$. If one wants to find out if there is a CI from $A$ at time
$t_0$ to $B$ at time $t$ or   from $B$ at time $t_0$ to $A$ at time
$t$, the initial state created by a do-intervention on either side
factors completely,  $\rho(t_0)=\rho^A\otimes \rho^B\otimes\rho^{C}$,
which is the form we assume from now on. 
 \DB{The initially factoring states avoids the problem that for an initially entangled state a measurement by Bob collapses the state also on Alice's side and would signal CI, while it is well known that in this scenario no information can be transmitted. }
  When we are    not concerned with time dependence we may skip the arguments $t$,
    $t_0$ and simply consider initial and final states of a joint
    evolution.
Def.~\ref{DefCausalInfluence} of no CI is similar to the one of a ``semicausal map'' in \cite{schumacher_locality_2005}, but is based on the dependence of a final state directly on an initial state rather than a local channel.  {The generalization of Def.~\ref{DefCausalInfluence}, as well as of Theorem \ref{Theorem no-causal influence} and expression (\ref{eq indicator simplifyed}) below to  $N$ quantum systems is given in the Supplemental Material (SM)}.

For  finite $d_C$,
we consider the 
evolution of $A,B,C$ via a joint
unitary transformation $U\in\mathcal{U}(d_Ad_Bd_C)$.
The uncorrelated initial state is mapped to
$U(\rho^A\otimes\rho^B\otimes\rho^C)U^{\dagger}$. 
{Using 
  from now on
  Einstein's sum convention,} we write the 
initial states as 
$
\rho^A=\rho_{ij}^A\ketbra{i}{j}$ and
$\rho^B=\rho_{kl}^B\ketbra{k}{l}$, for $i,j\in[d_A]$, $k,l\in[d_B]$,
and take $\rho^C=\ketbra{0}{0}_C$. Then, Alice's final reduced state is
\begin{equation}
\label{eq reduced density matrix A components}
    \rho'^A_{i'j'}=\rho_{ij}^A\rho_{kl}^BU_{i'k'm',ik0}U_{j'k'm',jl0}^{*},
\end{equation}
 where the prime indices run over the same range as the respective non-primed and $m'\in[d_C]$ and thus we see that $B\nrightarrow A$ if (\ref{eq reduced density
  matrix A components}) does not depend on $\rho^B$.

\begin{theorem}\label{Theorem no-causal influence} {Let $A$ and
    $B$ be quantum systems in a common environment $C$ with initial state $\ket{0}_C$
    and let $F_{kl,ij}^U(i',j'):=U_{i'k'm',ik0}U_{j'k'm',jl0}^{*}$. Then,
  $B\nrightarrow A$ if and only if, for all $i,j,k,\Tilde{k},l,i',j'$, (no sum over $\Tilde{k}$)}
    \begin{equation}\label{Eq no-causal influence condition}
        F_{kl,ij}^U(i',j')=\delta_{kl}F_{\Tilde{k}\Tilde{k},ij}^U(i',j').
    \end{equation} 
\end{theorem}  The statement for $A\nrightarrow B$ is analogous with
the function
$\Tilde{F}^{U}_{ij,kl}(k',l')=U_{i'k'm',ik0}U_{i'l'm',jl0}^{*}$
replacing $F_{kl,ij}^U(i',j')$. {On the other hand, if the evolution of the initial uncorrelated state is given by Kraus operators $K^{\mu}$, $\rho^{'ABC}=K^\mu\rho^{ABC}K^{\mu\dagger}$, Theorem \ref{Theorem no-causal influence} holds with $F^{K}_{kl,ij}(i',j'):=K^{\mu}_{i'k'm',ik0}K_{j'k'm',jl0}^{\mu*}$.} The proof of Theorem \ref{Theorem no-causal influence} \DB{is based on straight forward linear algebra and} is given in the SM.

\ssection{Mutuality}
As an example, consider $U=U^A\otimes
U^{BC}$, for $U^{A}\in\mathcal{U}(d_A)$, $U^{BC}\in\mathcal{U}(d_Bd_C)$, the unitary group. 
Then $F^U_{kl,ij}(i',j')$ becomes $\delta_{kl}U^A_{i'i}U^{A*}_{j'j}$, such that condition (\ref{Eq no-causal influence condition}) is fulfilled, and therefore, by Theorem \ref{Theorem no-causal influence}, $B\nrightarrow A$ 
as expected. 
Similarly one finds $A\nrightarrow B$.
However, \DB{tensor products of unitary matrices} are  a set of Haar  measure zero in
$\mathcal{U}(d_Ad_Bd_C)$. 
Unitary matrices that allow one-way CI exist as well. An
example where  $A\substack{\nleftarrow\\[-1em]\rightarrow } B$,
{
  is given by the
unitary transformation 
$U^{(123)}$  
corresponding to the 
permutation $
\ket{ikm}\mapsto\ket{mik}$. 
The opposite case 
results from
the
unitary 
$U^{(132)}$ that 
permutes
$
\ket{ikm}\mapsto\ket{kmi}$
}. See SM for another example
which does not correspond to a permutation. Among $10^5$ Haar {distributed} random unitary $2^3 \times
2^3$ matrices via QuTip \cite{JOHANSSON20131234}
all of them permitted  CI in both directions, 
which is the generic situation.

\ssection{No transitivity}
Let $A,B$ and $C$ be quantum systems and
$t_0<t_1<t_2$. $A(t_0)\rightarrow B(t_1)$ and $B(t_1)\rightarrow
C(t_2)$ does not imply $A(t_0)\rightarrow C(t_2)$. As a
counterexample, consider that at $t_1$ one applies the $CNOT$ gate
with $B$ the control  and $A$ the target to the initial state
$\rho^A\otimes\rho^B\otimes\rho^C$.  The reduced density matrix of
$B$, $\rho'^B(t_1)$,  depends on $\rho^A$ only in the off-diagonal
elements (see SM).  Similarly, 
starting at time $t_1$ with a product state
$\rho^A\otimes\rho^B\otimes\rho^C$, if one   applies at $t_2$  the
$CNOT$ gate with $C$ the target 
system and $B$ the control, one easily shows $B(t_1)\rightarrow
C(t_2)$, but the reduced density matrix of system $C$ only depends on
the diagonal components of  $\rho'^B$, which do not depend on $\rho^A$
after the first $CNOT$, and therefore $A(t_0)\nrightarrow C(t_2)$. 
However, one can easily construct corresponding examples in the
classical case.  Hence, CI must be distinguished from
implications as these \textit{are} transitive. 

\ssection{Measure of the causal influence} Consider the two qubit systems $A$ (control) and $B$ (target) and apply the $CNOT$ gate on a pure state. 
One might suspect that $CNOT$ permits only influence from
$A$ to $B$. However, applying Theorem \ref{Theorem no-causal
  influence}, we see that both CI directions are allowed\DB{. This is not merely a mathematical consequence of the definition but corresponds to a  well-known and real physical effect called \textit{phase kickback } in quantum circuits \cite{PhaseKickback} where the phase $\beta$ in the initial state $(\ket{0}+\ket{1})\otimes(\ket{0}+e^{i\beta}\ket{1})/2$, 
ends     up in the state of Alice. More generally, in the SM we give the reduced state after the $CNOT$ gate for both parties for an arbitrary initial product state and} one sees that in both cases these depend on components of their partner's initial state.
Nevertheless, while Alice's final state only carries Bob's dependence in the off-diagonal
components (see (C.1) in the SM), all his components after the $CNOT$ depend on Alice's
initial state.  This is in sync with the intuition that the CI from 
control to target is  'stronger' than 
the other way round,
and motivates
the introduction of a measure of CI. Although one could argue that  a $CNOT$ with reversed roles of control and target is a $CNOT$ conjugated with local Hadamard transformations and hence expect equal influence in both directions, we prove below that it is reasonable not to request invariance of a measure of CI under  local pre-propagation on the influencing system, as varying its initial state is part of the process of examining the influence. A natural way to quantify the CI is via  
$\abs{\partial \rho'^A_{i'j'}/\partial \rho^B_{hf}}$, for
$h,f\in[d_B]$. We base our definition on \DB{all} pure initial states $\rho^A$
of Alice.  
Note that due to linear propagation the
$\rho'^A_{i'j'}$ are holomorphic functions of the $\rho^B_{hf}$ such
that these complex derivatives
are always well-defined. For the same reason, the measure is independent of $\rho^B$. 
\begin{definition} \label{I_ABnonUdependentDef}
  Let $\rho^A=V\ketbra{0}{0}V^{\dagger}$, for $V\in\mathcal{U}(d_A)$ and
some fixed initial state $\ket{0}_A$ of Alice,
i.e. $\rho^A_{ij}=V_{i0}V_{j0}^*$. We define the measure of causal
influence
from $B$ to $A$ as 
\begin{equation}
\label{eq def CI}
    I_{B\rightarrow A}=\int d\mu(V)\sum_{hfi'j'}\abs{\partial
      \rho'^A_{i'j'}/\partial \rho^B_{hf}}^2. 
\end{equation}
\end{definition}
Analogously one defines $I_{A\rightarrow B}$.  Notice that for either $A$ or $B$ \DB{corresponding to} the trivial (1-dimensional) Hilbert space, there is a single physical state so the derivatives
vanish and the influence is always 0. {Therefore we restrict ourselves to $d_A,d_B>1$}. 
Based on an average rather than a maximization procedure, 
definition \ref{I_ABnonUdependentDef} allows for straight forward evaluation. While not all do-interventions are physically realizable \cite{Milburn2018}, both definitions \ref{DefCausalInfluence},\ref{I_ABnonUdependentDef} can be applied experimentally, as long as Alice and Bob can generate arbitrary local states, e.g.~through measurement and local quantum channels.  

\begin{figure}[h]
    \centering
    \subfigure[]{\includegraphics[width=30mm]{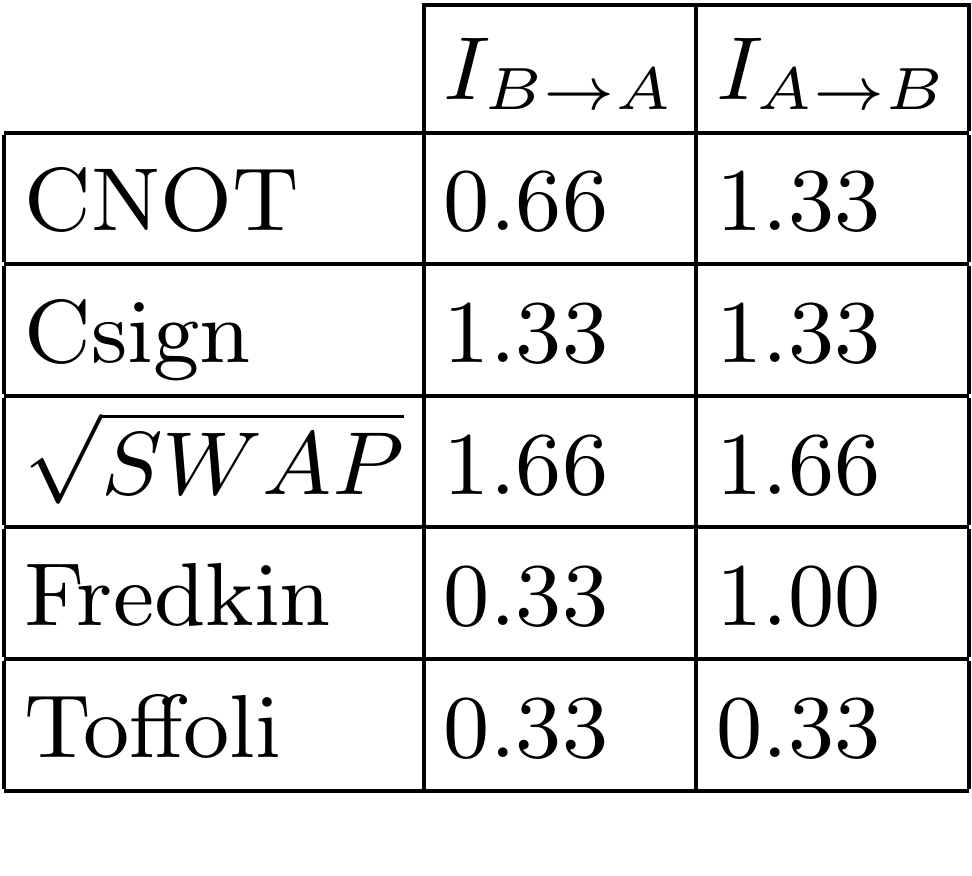}}
    \subfigure[]{\includegraphics[width=50mm]{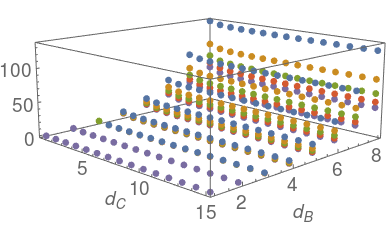}}
    \caption{(Color online) (a) CI for some quantum gates. The environment for the three-qubit gates is initially in state {$\ket{0}_C$} and, when required, $A$ is the control qubit and $B$, the target. (b) $E[I_{A\rightarrow B}]$ as function of $d_A,d_B$ and $d_C$ (eq.\eqref{eq EIU}), 
      where $d_A$ is coded in the sequence of points for each $d_B,d_C$ with $2\le d_A\le 6$ from top to bottom 
      ($3\le d_B\le 8$) or bottom to top ($d_B=2$, co-inciding on this scale) in each sequence. 
    }
    \label{fig.EIab}
\end{figure}

For finite $d_C$, from (\ref{eq reduced density matrix A components}) and the definition of $F$, $\partial \rho'^A_{i'j'}/\partial \rho^B_{hf}=\rho^A_{ij}
F_{kl,ij}^U(i',j')D_{kl,hf}$ with $D_{kl,hf}\equiv \partial\rho_{kl}^B/\partial\rho_{hf}^B$. Since for all density matrices $\rho$, $\rho_{i_0i_0}=1-\sum_{j\neq
  i_0}\rho_{jj}$ and the partial derivative of a complex number with
respect to its complex conjugate vanishes, we have
$D_{kl,hf}=(1-\delta_{hf})\delta_{kh}\delta_{lf}-\delta_{hf}\delta_{kl}(\delta_{h0}-1)(-1)^{\delta_{k0}}$,
where, without loss of generality, we took $i_0=0$. Then, \DB{by algebraic computation and using invariant integration over the unitary group   \cite{Aubert2003}, see SM,}
\begin{equation}  \label{eq indicator simplifyed}
\begin{split}
     &I_{B\rightarrow A}(U)=\big[
     F_{kl,ij}^U(i',j')F_{k_1l_1,ij}^{*U}(i',j')\\
     &+ F_{kl,ii}^U(i',j')F_{k_1l_1,i_1i_1}^{*U}(i',j')\big]\frac{D_{kl,hf}D_{k_1l_1,hf}}{d_A(d_A+1)}
     .   
\end{split}
\end{equation}
{If the evolution is given by Kraus operators $K^{\mu}$, the 
  CI is computed replacing $F^U$ 
  by $F^{K}$}. {In} Fig. \ref{fig.EIab}(a) {we show} the values of the 
CI for some common qubit gates. 

\begin{figure}[h]
\centering
\includegraphics[width=80mm]{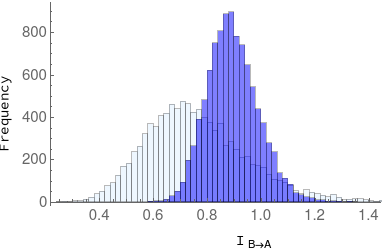}
\caption{(Color online)  Histogram of $I_{B \rightarrow A}$, for $10^4$ random Haar distributed unitary matrices of dimension $D=d_Ad_Bd_C$, for $D=2\cdot2\cdot2$ (light blue) and $D=3\cdot2\cdot2$ (dark blue), their mean values are 0.76 and 0.896, the standard deviations 0.19 and 0.095, and their expected values (eq.\eqref{eq EIU}) $16/21\simeq 0.76$ and $128/143\simeq 0.895$, respectively.} 
\label{Figure E[I(U)] fixed dC}
\end{figure}

\ssection{Properties of the measure}{The measure (\ref{eq def CI}) enjoys the following important properties (see SM for the proofs):}\\
1. $I=0$ if and only if there is no CI.\\
2. $I_{B\rightarrow A}((U_1^A\otimes U^B\otimes\mathbb{I}^{C}) U (U_0^A\otimes \mathbb{I}^{BC}))=I_{B\rightarrow A}(U),$ for all $U_0^A,U_1^A\in\mathcal{U}(d_A)$, $U^B\in\mathcal{U}(d_A)$ $U\in\mathcal{U}(d_Ad_Bd_C)$, and analogously for $I_{A\rightarrow B}$. These properties are natural: after the propagation, any local action should not change the CI. And since we consider all initial states of Alice, the measure should be invariant under her local pre-propagation unitaries. 
On the other hand, since the changes of $\rho^B$ reflect Bob's do-interventions, $I_{B\rightarrow A}$ need not be invariant under local unitaries of Bob before the propagation, and indeed, in general $ I_{B\rightarrow A}( U (\mathbb{I}^{AC} \otimes U^B))\ne I_{B\rightarrow A}(U)$.\\
3. The natural scale of $I_{B \rightarrow A}$ is given by the Haar-measure average $E[I_{B
  \rightarrow A}(U)]=\int d\mu(U) I_{B \rightarrow A}(U)$. 
\begin{equation}\label{eq EIU}
\begin{split}
       E[I_{B \rightarrow A}(U)]&=
       \frac{d_B-1}{d_A^2 d_B^2 d_C^2-1} [2 d_A^2 d_B^2 d_C-2 d_B^2 d_C  \\&+  d_A (d_B-2)^2 (d_B^2 d_C^2-1)].
\end{split}
\end{equation}
$E[I_{A \rightarrow B}(U)]$ is obtained by permuting $A\leftrightarrow B$.
  Fig.~\ref{fig.EIab}\DB{(b)} 
  illustrates the behaviour of $E[I_{A \rightarrow B}(U)]$. Notice that it 
is strictly increasing with $d_B$, which is 
reasonable as one expects that the larger the Hilbert space dimension of the influencing system, the more it can influence.
A particular
case of interest is such that either one of the systems $A$ or $B$, or
the environment has a dimension much greater than the others.  
  $E[I_{B \rightarrow A}(U)]$ tends to
$2(d_B-1)/d_C$, $\infty$ and $(d_B-1)(d_B-2)^2/d_A$ in these limits, respectively. 
Fig. \ref{Figure E[I(U)] fixed dC} shows a
histogram of $I_{A \rightarrow B}$ for random Haar generated unitary
matrices. %
The relatively narrow distribution confirms that $E[I_{B \rightarrow A}(U)]$ represents a good scale for $I_{B \rightarrow A}(U)$ for given Hilbert space dimension. For three qubits, there are gates such as CNOT or Csign with almost twice the value of $E[I_{B \rightarrow A}(U)]$. 

\ssection{Quantum causal switch}\DB{In analogy to the quantum switch \cite{Chiribella2012, Chiribella2013} one can create a unitary evolution $U_{sup}$ 
  that superposes the 
  CI $A\substack{\nleftarrow\\[-1em] \rightarrow} B$ and $A\substack{\leftarrow\\[-1em] \nrightarrow} B$, 
  controlled by an ancilla qubit $\ket{\chi^c}=\cos(\theta/2)\ket{0}+e^{i\phi}\sin(\theta/2)\ket{1}$, for $\theta\in[0,2\pi)$ and $\phi\in[0,\pi)$.  In the SM  we investigate the behavior of such a "quantum causal switch" and show how one can interpolate continuously between the two directions of the CI as function of the state of the control qubit. For almost all states of the control qubit, the CI is in both directions.}
  
\ssection{Propagation of   CI } \DB{In order to illustrate the propagation of CI within an exactly solvable physical model, compare it to the well-known creation of reservoir-induced entanglement, and to verify that indeed entanglement does not arrive before the CI, which would be physically unreasonable, we apply the measure to two spins-1/2 interacting with a bath of harmonic oscillators, which corresponds to $d_C=\infty$}. The Hamiltonian is $H=H_{AB}+H_{bath}+H_{int}$, with $H_{int}=(S^A+S^B)\DB{\sum_kg_kq_k}$. $S^A$, $S^B$ are the "coupling agents" acting on $\mathcal{H}_A$ and $\mathcal{H}_B$ with eigenvalues
$a_0,a_1$ and $b_0,b_1$, respectively, and
$q_k$ and $g_k$ are the generalized coordinates and coupling
constants to the $k$th oscillator, respectively.  For degenerate-in-energy, non-interacting spins $H_{AB}=0$, and the model becomes an exactly solvable dephasing model that can lead to reservoir-induced entanglement  
\cite{Braun2001}. \DB{The Hamiltonian does not give any direct interaction between $A$
      and $B$, but an effective interaction is mediated by the heatbath
      that turns out to be the standard dipole-dipole interaction with
      a time-dependence reflecting retardation.}
\DB{The propagation of an initial product state at time $t$ \cite{Braun2001} leads to a CI}
\begin{equation}\label{eq I heat bath}I_{B \rightarrow A}^{bath}=\frac{4}{3}\sin^2[(a_0-a_1)(b_0-b_1)\varphi(t)]e^{-2(a_0-a_1)^2f(t)},\end{equation}
\DB{where $\varphi(t)=\sum_k\frac{g_k^2}{2m\hbar\omega_k^2}(t-\frac{\sin\omega_kt}{\omega_k})$ and $f(t)=\sum_k\frac{g_k^2(1+2\Bar{n}_k)}{2m\hbar\omega_k^3}(1-\cos\omega_kt)$, where $\Bar{{n}}_k$ is the thermal occupation of the $k$th oscillator.}
{ Both  $\varphi(t)$ and $f(t)$} vanish at $t=0$ and {are} strictly
positive for $t>0$. 

$I_{A \rightarrow B}^{bath}$ is obtained by permuting $a_i\leftrightarrow
b_i$, $i=0,1$. 
The CI is \DB{ invariant under $t\to -t$}. 
For values of $f$ and $\varphi$ such that there is no causal
influence,
{no} entanglement between the qubits { is generated}. {However, the converse is not true,
  i.e.~not all 
  CI generates entanglement}. {$I_{B
    \rightarrow A}^{bath}$ takes} a maximum {value} of $4/3$,
periodically {oscillates as function of} $\varphi$, and 
decays exponentially for 
large $f$ \DB{ but does not vanish exactly, whereas the generated entanglement does vanish exactly} as the state approaches the fully mixed state \cite{PhysRevA.63.032307,PhysRevA.64.012316,PhysRevA.95.012318}. 

\begin{figure}[h]
\centering
 \subfigure[]{\includegraphics[width=40mm]{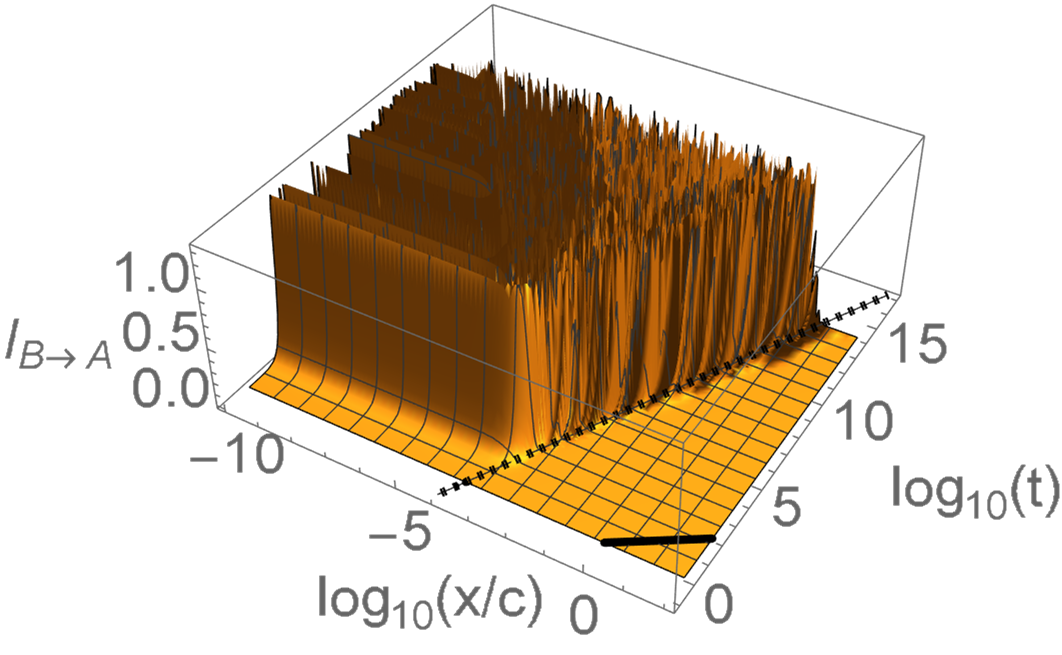}}
\subfigure[]{\includegraphics[width=40mm]{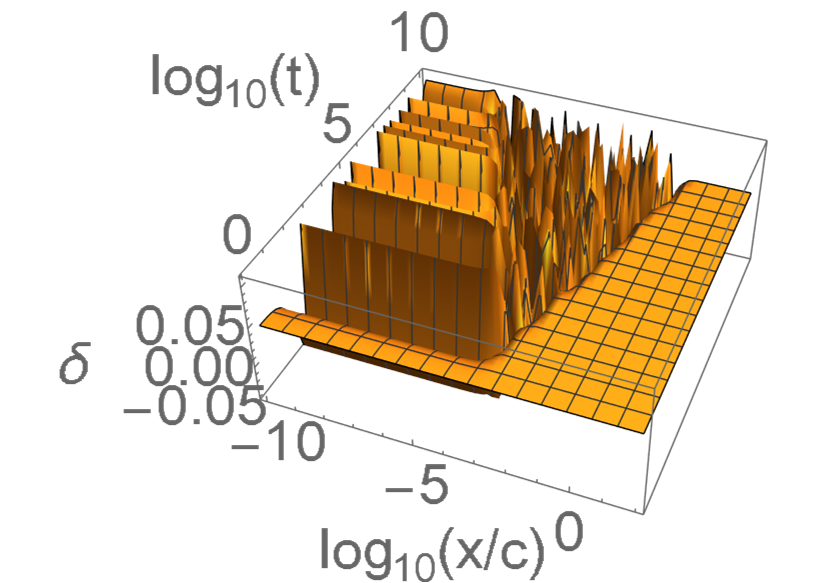}}
\caption{ (Color online)
  (a)
  ${I_{B\rightarrow A}}$ for two initially non-interacting DQD with $d=10$\,nm coupled
  to the black body radiation at $T=2.73K$ with $y_m=4250$ ($\hbar\omega_m=1$\,eV), parametrized by \DB{$x/c$} and $t$. The corresponding plot for the entanglement of formation  looks almost identical (see Fig.2 in \cite{Braun2002}). Dotted line: 
  $t=  10^{11}(\DB{x/c})^{3}$.
  Full line: light-cone $ct=x$.
 (b) Difference $\delta ={0.795481 I_{B\rightarrow A}}-E$, see text,   $E$ for the initial state $(\ket{0}+\ket{1})\otimes(\ket{0}+\ket{1})/2$.  Same parameters as in (a). }  
\label{Figure causal influence CMB}
\end{figure}

{For the physical example
  of two double quantum dots (DQD) at distance \DB{$x$} from each other coupled with dipole interaction to
  black-body radiation, with a UV frequency cut-off $y_m=\omega_{max}\tau $, where $\tau=\hbar/k_BT$ and $T$ is the temperature,
the functions $f(t)$ and $\varphi(t)$ can be {obtained} analytically (see SM) if one approximates coth $\simeq1$ for $y_m\gg1$ \cite{Braun2005}}. \DB{Physically, it is clear that CI can only arise inside the light-cone, $x=c t$, and indeed, the CI plotted for this system in Fig. \ref{Figure causal influence
  CMB}{(a)} shows that there is no 
CI for spacelike separated points, $t<x/c$}.
Surprisingly, however, significant CI is generated only
far behind the light-cone, namely for
$t\gtrsim 10^{12}\DB{(x/c)}^3$. This is reminiscent of reservoir-induced entanglement \cite{Braun2002,benatti_environment_2003}
that also arises only far behind the light-cone
\cite{Braun2005} (i.e.~``Entanglement harvesting'' \cite{Reznik2005,Ball2006,Lin2010,Olson2011,Olson2012,MartnMartnez2014,ralph2014quantum,MartnMartnez2016,PozasKerstjens2016,Ng2018,Xu2020}, i.e.~entanglement creation outside the light-cone,  is typically not possible here without ``reservoir engineering'').  
In fact, the space-time dependence of CI and entanglement of formation (EOF) $E$, created as long as both initial states of $A$ and $B$ contain components of both $\ket{0}$ and $\ket{1}$, are almost perfectly in sync, see Fig.\ref{Figure causal influence CMB}(b). Minimization of $\sum_{i_t,\DB{j_{x}}} \delta^2_{i_t,\DB{j_{x}}}$ over a regular grid of 51$\times$51 points in the space-time regions shown with  $\delta ={\lambda I_{B\rightarrow A}}-E$ gives  $\lambda\simeq 0.795481$, and the remaining differences are less than about $0.05$ in absolute value over the 12 orders of magnitude of \DB{$x/c$} considered. 
Under time reversal, $t\to -t$, $f(t)$ remains invariant, whereas
$\varphi(-t)=-\varphi(t)$, which leads to complex conjugate matrix
elements of $\rho^{AB}$ as usual. Formally 
there is hence exactly as
much 
CI in positive time direction as in negative time
direction.  This can be seen already from \eqref{I_ABnonUdependentDef} 
with $\rho'^{AB}\to \rho'^{AB*}$ under time-reversal.
Hence, reasons for the apparent purely forward 
CI in Nature must be sought outside quantum mechanics \cite{purves2021quantum}.  Indeed, causality  
is, even in our most fundamental established theories, implemented by
hand by choosing advanced Green's functions only.  

{\em In summary,} we gave a definition of causal influence (CI) 
  in the quantum world based on reduced density matrices. We derived a
  necessary and sufficient condition \DB{on the joint evolution operator of Alice, Bob, and an environment}
  that a given quantum system
  can causally influence another one. Moreover, we introduced a
  measure of the CI, analysed it in detail, 
  {showed the possibility of superposing 
    opposite directions of CI, 
  } and applied
  it to particular cases of both finite and infinite dimensional
  environments.  For the 
  example of two degenerate double-quantum
  dots \DB{at distance $x$} dipole-interacting with thermal black body radiation we found
  that the space-time dependence of the CI is almost perfectly in sync with the reservoir induced entanglement.  Both arrive long after
  ($t\propto \DB{(x/c)}^{3}$)
  the light cone $ct=x$.
  \DB{Just as entanglement measures, classicality measures and measures
based on other resource theories have had a large impact in
theoretical physics for the last three decades.  Similarly, we hope that
having a causality measure in quantum mechanics opens the path to study many new things, starting from its properties, over Bell-like 
inequalities, to the relationship to entanglement harvesting, the classical limit of quantum CI, and many more.}

\ssection{Acknowledgements} We thank Lukas Fiderer and Kilian Stahl
for discussions, Ronny M\"uller for suggesting the unitary in eq.(2) 
in the SM for a one-way CI and Marten Folkertsma for discussions about invariances of the measure. 
\bibliographystyle{apsrev4-1}

\bibliography{references_paper}

\begin{widetext}
\newpage
\begin{center} \bf SUPPLEMENTARY MATERIAL\end{center}

\section{A. Proof of Theorem 1}

Let $\alpha_{ij}(i',j'):=\rho_{kl}^BU_{i'k'm',ik0}U_{j'k'm',jl0}^{*}$. Then
from (1), 
 $\rho'^A_{i'j'}=\rho_{ij}^A\alpha_{ij}(i',j')$. Imposing
$B\nrightarrow A$ for all $\rho^A$,  $B\nrightarrow A$ must hold in
particular  for $\rho^A_{ij}=\delta_{ij}\delta_{ii_0}$ and thus,
$\rho'^{A}_{i'j'}=\alpha_{i_0i_0}(i',j')$ must be independent of
$(\rho^B)\,\,\forall i_0$. Since $i_0$ is arbitrary $\in [d_A]$, we
need that  $\alpha_{ii}(i',j')(\slashed{\rho}^B) \forall i\in [d_A]$\DB{, where $(\slashed{\rho}^B)$ denotes no dependence on $\rho^B$ for all $\rho^B$}.
Moreover, consider the previous density matrix but with two
non-vanishing off-diagonal components $\rho^A_{i_1j_1}$ and, due to
Hermiticity, $\rho^A_{j_1i_1}$, for $i_1\neq j_1$. Then, imposing the
independence condition, it is necessary that
\begin{equation}
   \big(\rho_{i_1j_1}^A\alpha_{i_1j_1}(i',j')+\rho_{j_1i_1}^A\alpha_{j_1i_1}(i',j')\big)(\slashed{\rho}^B) 
\end{equation}
(without implicit sum). Thus, writing $\rho_{i_1j_1}^A$ as the sum of
its real part and $i$ times its imaginary part and $\rho_{j_1i_1}^A$
as {their} difference, one sees that it is necessary that
$(\alpha_{i_1j_1}(i',j')\pm\alpha_{j_1i_1}(i',j'))(\slashed{\rho}^B)$,
leading to the conclusion that
{$\alpha_{ij}(i',j')(\slashed{\rho}^B)$ must hold for all
  $i,j,i',j'$.} Applying the definition of $F_{kl,ij}^U(i',j')$,
  \begin{equation}
  \label{eq alpha_ij splitted}
     \alpha_{ij}(i',j')=\rho_{kk}^BF_{kk,ij}^U(i',j')+\sum_{l\neq
  k}\rho_{kl}^BF_{kl,ij}^U(i',j'),
  \end{equation}
 its independence of $\rho^B$ is
fulfilled if and only if 
\begin{eqnarray}
  \label{eq:cond1}
F_{kk,ij}^U(i',j')&=&F_{\Tilde{k}\Tilde{k},ij}^U(i',j') \mbox{ and }\nonumber\\
F_{kl,ij}^U(i',j')&=&F_{lk,ij}^U(i',j')=0, 
\end{eqnarray}
for all $k\neq l,\Tilde{k},i,j,i',j'$.  The first condition comes from
the trace one of the density matrices, thus, any $\rho^B_{k_0k_0}$ can
be written as $\rho^B_{k_0k_0}=1-\sum_{k\neq k_0}\rho^B_{kk}$ and thus
the first summand in (\ref{eq alpha_ij splitted}) becomes
\begin{equation}
    \sum_{k\neq
  k_0}\rho_{kk}^BF_{kk,ij}^U(i',j')+(1-\sum_{k\neq
  k_0}\rho^B_{kk})F_{k_0k_0,ij}^U(i',j').
\end{equation} 
Since $\rho_{kk}^B$ for $k\neq k_0$ can vary independently, it is
necessary that $F_{kk,ij}^U(i',j')=F_{k_0k_0,ij}^U(i',j')$ for all
$k$, then
\begin{equation}
    \sum_k\rho_{kk}^BF_{kk,ij}^U(i',j')=F_{k_0k_0,ij}^U(i',j')\sum_k\rho_{kk}^B
=F_{k_0k_0,ij}^U(i',j')\tr{\rho^B}=F_{k_0k_0,ij}^U(i',j').
\end{equation}
Finally,   
the second condition comes from a similar argument derived for the
independence of $\alpha_{ij}(i',j')$ with $i\neq j$.

\section{B. Example of a single direction of influence}
The unitary matrix (\ref{eq unitary only one direction}) in $\mathcal{U}(2\cdot2\cdot2)$ is such that $A\rightarrow B$ but $B \nrightarrow A$, i.e. it only allows one influencing direction. Moreover, $I_{A\rightarrow B}(U)=1.5$ and $I_{B\rightarrow A}(U)=0.0$.
\begin{equation}
\label{eq unitary only one direction}
    U=\begin{pmatrix} 1&0&0&0&0&0&0&0  \\
                      0&0&1&0&0&0&0&0\\
                      0&0&0&0&\frac{1}{\sqrt{2}}&\frac{1}{\sqrt{2}}&0&0\\
                      0&0&0&0&0&0&\frac{1}{\sqrt{2}}&\frac{1}{\sqrt{2}}\\
                      0&0&0&0&\frac{1}{\sqrt{2}}&-\frac{1}{\sqrt{2}}&0&0\\
                      0&0&0&0&0&0&\frac{1}{\sqrt{2}}&-\frac{1}{\sqrt{2}}\\
                      0&1&0&0&0&0&0&0 \\
                      0&0&0&1&0&0&0&0 
    
    \end{pmatrix}.
\end{equation}

\section{C. Qubit density matrices after the application of the CNOT gate}
Writing Alice's and Bob's initial states in the computational basis so that $ \rho^A=\rho_{ij}^A\ketbra{i}{j}$ and $\rho^B=\rho_{kl}^B\ketbra{k}{l}$, for $i,j\in[2]$, $k,l\in[2]$,  the joint state after the application of the $CNOT$ gate to the state $\rho^A\otimes\rho^B$ is given by $CNOT(\rho^A\otimes\rho^B)(CNOT)^{\dagger}$ and, tracing out systems $B$ and $A$, respectively, one has
\begin{equation}
\label{eq states after CNOT}
    \rho'^A=\begin{pmatrix}  \rho_{00}^A&\rho_{01}^A2\operatorname{Re}\rho_{01}^B \\\rho_{10}^A2\operatorname{Re}\rho_{10}^B & \rho_{11}^A\end{pmatrix}, \hspace{5mm} \rho'^B=\begin{pmatrix} \rho_{00}^B\rho_{00}^A+\rho_{11}^B\rho_{11}^A &  \rho_{01}^B\rho_{00}^A+\rho_{10}^B\rho_{11}^A\\ \rho_{01}^B\rho_{11}^A+\rho_{10}^B\rho_{00}^A & \rho_{00}^B\rho_{11}^A+\rho_{11}^B\rho_{00}^A\end{pmatrix}.
\end{equation}

\section{D. Integration over the Unitary Group}
Using the definition of $D_{kl,hf}$,  
the CI (3) can be written as
\begin{equation}
    \label{I_AB Haar def}
    I_{B\rightarrow A}(U)=\int d\mu(V)\sum_{hfi'j'}\abs{V_{i0}V_{j0}^*
      F_{kl,ij}^U(i',j')D_{kl,hf}}^2. 
\end{equation}
Expanding the square, one has order $2$ terms of the functions $F$ and
$D$ and integrals of the type $\int d\mu(V)V_{j0}^*V_{i_10}^* V_{i0}
V_{j_10}$.

In [50] 
is shown that the integrals $\int d\mu(U)U_{i_1j_1}^*...U_{i_pj_p}^*U_{k_1l_1}...U_{k_ql_q}$ denoted by $\braket{I,J}{K,L}$ vanish unless $q=p$, which will be assumed to be the case. Even more, such integrals are non-vanishing if, in addition, $K=I_{\sigma_1}$ and $L=J_{\sigma_2}$, for $\sigma_1,\sigma_2\in S_p$, and since $U_{i_1j_1}...U_{i_pj_p}=U_{\tau(i_1)\tau(j_1)}...U_{\tau(i_p)\tau(j_p)}$ for any $\tau\in S_p$ due to the fact that it is a multiplication of complex numbers, we may assume $\sigma_1=id$, then the non zero integrals are of the form $\braket{I,J}{I,J_{\sigma_2}}$. Since the Haar measure is invariant under transposition, a permutation between rows and columns does not change the integral, i.e. $\braket{I,J}{K,L}=\braket{J,I}{L,K}$. Furthermore, the integral is affected by whether the indices take on the same or different values, but independent of what these values are, for this reason {it} is convenient to use the graphical representation introduced in [50]:
\begin{enumerate}
    \item The distinct values in the index set $I$ are represented as dots in a column and, on its right, the distinct values of the index set $J$ as dots in a column (since $J_{\sigma_2}$ is a permutation of $J$, it has the same distinct values {as $J$}).
    \item Factors $U^*_{i_rj_r}$ and $U_{i_r,\sigma_2(j_r)}$, for $r=1,...,p$, are represented by thin (solid) and dotted lines, respectively. The power of the matrix element, if grater than 1, will be represented above the solid line or below the dotted line, correspondingly. When a pair $U^*_{i_rj_r}U_{i_rj_r}$ occurs together, the thin and dotted lines will be replaced by a thick solid line, whose multiplicity will be understood as the power of this pair.
\end{enumerate}
Using this graphical representation, the non-vanishing integrals { $
  \mathcal{I}$
} of $p=2$  are those shown in Figure \ref{Figure order 2 integrals} together with their values. Then, the non-vanishing integrals $\int d\mu(V)V_{j0}^*V_{i_10}^* V_{i0} V_{j_10}=\braket{ji_1,00}{ij_1,00}$, for $V\in\mathcal{U}(d_A)$, are such that $(i,j_1)=\sigma(j,i_1)$, for $\sigma\in S_2$, i.e. either
\begin{enumerate}
    \item $(i,j_1)=(j,i_1)$, so that the integral is $\braket{ii_1,00}{ii_i,00}$, and it takes the values
    \begin{equation}
         \braket{ji_1,00}{ji_i,00}=\begin{cases}
               \mathcal{I}_{(e)}=\frac{2}{d_A(d_A+1)} & \text{if }j=i_1,\\
               \mathcal{I}_{(d)}=\frac{1}{d_A(d_A+1)} & \text{if }j\neq i_1,
           \end{cases}
    \end{equation}
   whose value can be compactly written as $\frac{1+\delta_{i_1j}}{d_A(d_A+1)}$,
    \item or $(i,j_1)=(i_1,j)$, and, since the integral is independent of what the values of the indices are, it takes the same value as in case 1.
\end{enumerate}
Thus, considering both options and not overcounting the intersection, 
\begin{equation}
\label{eq explicit integral VdA}
\begin{split}
      \int d\mu(V)V_{j0}^*V_{i_10}^* V_{i0} V_{j_10}&=
       \delta_{i_1i}\delta_{jj_1}[\delta_{ji_1}\mathcal{I}_{(e)}+(1-\delta_{ji_1})\mathcal{I}_{(d)}]+
       \delta_{i_1j_1}\delta_{ij}[\delta_{ji_1}\mathcal{I}_{(e)}+(1-\delta_{ji_1})\mathcal{I}_{(d)}]-
       \delta_{i_1j}\delta_{i_1i}\delta_{i_1j_1}\mathcal{I}_{(e)}\\&=
       \frac{1}{d_A(d_A+1)}( \delta_{i_1i}\delta_{jj_1}+\delta_{i_1j_1}\delta_{ij}).
\end{split}
\end{equation}
\begin{figure}[h]
\centering
\subfigure[]{\includegraphics[width=50mm]{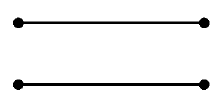}}
\subfigure[]{\includegraphics[width=50mm]{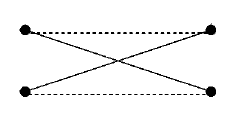}}
\subfigure[]{\includegraphics[width=50mm]{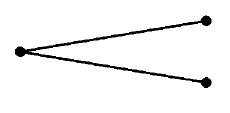}}
\subfigure[]{\includegraphics[width=50mm]{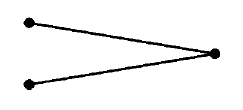}}
\subfigure[]{\includegraphics[width=50mm]{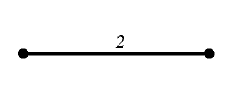}}
\caption{[50] Non-vanishing integrals { $
    \mathcal{I}$ 
  } for $p=2$. Their respective values, from (a) to (e), are: $\frac{1}{D^2-1}=:\mathcal{I}_{(a)}$, $\frac{-1}{D(D^2-1)}=:\mathcal{I}_{(b)}$, $\frac{1}{D(D+1)}=:\mathcal{I}_{(c)}$, $\frac{1}{D(D+1)}=:\mathcal{I}_{(d)}$ and $\frac{2}{D(D+1)}=:\mathcal{I}_{(a)}$.}
\label{Figure order 2 integrals}
\end{figure}

\section{E. Generalization to $n$ quantum systems}

Consider $n$ quantum systems described by their respective density matrices
$\rho^1$,...,$\rho^n$ and a system $e$, the joint environment,
with a fixed initial state $\rho^e$ corresponding to the physical
system that propagates the causal influence and creates an effective
interaction, but also leads to decoherence.  The joint quantum system
lies in the Hilbert space $\mathcal{H}_1\otimes...\otimes\mathcal{H}_n\otimes\mathcal{H}_e$, with
dim$(\mathcal{H}_r)=d_r$, $r\in\{1,...,n,e\}$. Definition 1 naturally extends as: $l\nrightarrow k$ if and only if, after the propagation of the initial state $\rho^1\otimes...\otimes\rho^n\otimes\rho^e$, the reduced state of system $k$ fulfils that $\rho'^{k}(\slashed{\rho}^l)$ for any density matrix $\rho^l$ describing system $l$.

Definition 2 is generalized as: Let $\rho^k=V^k\ketbra{0}{0}V^{k\dagger}$ for, where $V^k\in\mathcal{U}(d_k)$ so that $\rho^k_{i_kj_k}=V^k_{i_k0}(V_{j_k0}^{k})^*$. Then, the causal influence from $l$ to $k$ is
\begin{equation}
    I_{l\rightarrow k}=\int d\mu(V^k)\sum_{i'_kj'_k\Tilde{i_l}\Tilde{j_l}}\abs{\frac{\partial
      \rho'^k_{i_k'j_k'}}{\partial \rho^l_{\Tilde{i_l}\Tilde{j_l}}}}^2. 
\end{equation}

{If the full system is closed}, {we consider the evolution of the systems $1,...,n,e$ via a joint
  unitary transformation} $U\in\mathcal{U}(d_1\cdot...\cdot d_n \cdot d_e)$. 
The uncorrelated initial state is mapped to
$U(\rho^1\otimes...\otimes\rho^n\otimes\rho^e)U^{\dagger}=:\rho'^{1...ne}$. 
{Using Einstein's sum convention,} we write the
initial state of system $r$ in the computational basis so that $
\rho^r=\rho_{i_rj_r}^A\ketbra{i_r}{j_r}$, for $i_r,j_r\in[d_r]$ for all $r\in\{1,...,n\}$ and take $\rho^e=\ketbra{0}{0}$. Then,
\begin{equation}
    \rho'^{1...ne}_{i'_1...i'_ni'_e,j'_1...j'_nj'_e}=\rho^1_{i_1j_1}...\rho^n_{i_nj_n}U_{i'_1...i'_ni'_e,i_1...i_n0}U^*_{j'_1...j'_nj'_e,j_1...j_n0}.
\end{equation}
Then, the elements of the reduced density matrix of the system $k$ become
\begin{equation}
\label{eq reduced rho extended}
\rho'^k_{i'_k,i'_j}=\rho^1_{i_1j_1}...\rho^n_{i_nj_n}U_{i'_1...i'_{k-1}i'_ki'_{k+1}...i'_ni'_e,i_1...i_n0}U^*_{i'_1...i'_{k-1}j'_ki'_{k+1}...i'_ni'_e,j_1...j_n0}.
\end{equation}
Defining
\begin{equation}
\label{eq def F extended}
    F_{i_1j_1...i_nj_n}^{U}(i'_k,j'_k)=U_{i'_1...i'_{k-1}i'_ki'_{k+1}...i'_ni'_e,i_1...i_n0}U^*_{i'_1...i'_{k-1}j'_ki'_{k+1}...i'_n,i'_e,j_1...j_n0},
\end{equation}
 
and, from (\ref{eq reduced rho extended}) and (\ref{eq def F extended}), 
\begin{equation}
    \frac{\partial
      \rho'^k_{i_k'j_k'}}{\partial \rho^l_{\Tilde{i_l}\Tilde{j_l}}}=\prod_{r\neq l}\rho^r_{i_rj_r}F_{i_ij_1...i_nj_n}^{U}(i'_k,j'_k)D_{i_lj_l,\Tilde{i_l}\Tilde{j_l}},
\end{equation}
thus, 
\begin{equation}
     I_{l\rightarrow k}=F_{i_1j_1...i_nj_n}^{U}(i'_k,j'_k)
                        F_{i_{1_2}j_{1_2}...i_{n_2}j_{n_2}}^{*U}(i'_k,j'_k)D_{i_lj_l,\Tilde{i_l}\Tilde{j_l}}
                   D_{i_{l_2}j_{l_2},\Tilde{i_l}\Tilde{j_l}}\prod_{r,r_2\neq k,l}\rho^r_{i_r,j_r}\rho^{r_2*}_{i_{r_2},j_{r_2}}\int d\mu(V^k)\rho^k_{i_k,j_k}\rho^{k*}_{i_{k_2},j_{k_2}},
\end{equation}
where the values of the integral is given by (\ref{eq explicit integral VdA}).\\
On the other hand, applying an analogous reasoning {as} in the proof of Theorem 1, one finds the extension of Theorem 1: $l\nrightarrow k$ if and only if
\begin{equation}
\prod_{r\neq k,l}\rho^r_{i_rj_r}F_{i_1j_1...i_lj_l...i_nj_n}^{U}(i'_k,j'_k)=\delta_{i_lj_l}\left(\prod_{r\neq k,l}\rho^r_{i_rj_r}F_{i_1j_1...\Tilde{i_l}\Tilde{i_l}...i_nj_n}^{U}(i'_k,j'_k)\right),\end{equation}
for all $i_l,j_l,\Tilde{i_l},i_k,j_k,i_k',j_k'$.

\section{F. Proof of Property 1. of the measure: $I=0$ if and only if there is no causal influence}
By Def 1, if there is no causal influence, all the derivatives in (3) 
vanish and therefore $I$ is the integral over $0$, which is $0$. 
Conversely, suppose that $\partial\rho'^A_{i'j'}/\partial \rho^B_{hf}\ne0$ for a particular combination of indices $h,f$ and a specific value of $\rho^B_{hf}$.  Since $\rho'^A_{i'j'}$ is a linear function of all $\rho^B_{hf}$, $\partial\rho'^A_{i'j'}/\partial \rho^B_{hf}$ is independent of $\rho^B_{hf}$. 
Hence, $\partial\rho'^A_{i'j'}/\partial \rho^B_{hf}>0$ for a set of finite measure, and hence $I_{B\to A}\ne 0$.



\section{G. Proof of Property 2. of the measure: $I_{B\rightarrow A}((U_1^A\otimes U^B\otimes\mathbb{I}^{C}) U (U_0^A\otimes \mathbb{I}^{BC}))=I_{B\rightarrow A}(U).$}

The invariance follows from the following three invariances:
\begin{enumerate}
    \item  $I_{B\rightarrow A}((U^A\otimes \mathbb{I}^{BC}) U )=I_{B\rightarrow A}(U).$
    
    According to (4), 
     the dependence of the measure of the causal influence on the unitary relies on the sum of the functions $F$. Let $W:=(U^A\otimes\mathbb{I}^{BC})\cdot U$, and consider the sum 
\begin{equation}
\label{eq sum FwFw}
    \sum_{\substack{i,j,i',j'}}F_{kl,ij}^W(i',j')F_{k_1l_1,i,j}^{*W}(i',j')=
    \sum_{\substack{i,j,i',j'}}(\sum_{k',m'}W_{i'k'm',ik0}W_{j'k'm',jl0}^{*})(\sum_{\Tilde{k}',\Tilde{m}'}W_{i'\Tilde{k}',\Tilde{m}',ik_10}^{*}W_{j'\Tilde{k}',\Tilde{m}',jl_10}).
\end{equation}
The components of $W$ are given by $W_{i'k'm',ik0}=\sum_{r_1=0}^{d_A-1}U^A_{i',r_1}U_{r_1k'm',ik0}$,
in such a way that \eqref{eq sum FwFw} becomes
\begin{equation}
    \sum_{\substack{i,j,k',m'\\\Tilde{k}',\Tilde{m}'}}
    \left(\sum_{r_1,r_3}U_{r_1k'm',ik0}U^*_{r_3\Tilde{k}'\Tilde{m}',ik_10}\sum_{i'}U^A_{i',r_1}U^{*A}_{i',r_3}\right)\left(\sum_{r_2,r_4}U^*_{r_2k'm',jl0}U_{r_4\Tilde{k}'\Tilde{m}',jl_10}\sum_{j'}U^{*A}_{j',r_2}U^A_{j',r_4}\right),
\end{equation}
using that $U^A$ is a unitary matrix, we have that the sums over $i'$ and $j'$ are $\delta_{r_1,r_3}$ and $\delta_{r_2,r_4}$, respectively. Therefore, since $r_1$ and $r_2$ are mute indices, we can replace them by $i'$ and $j'$, respectively and,  with implicit sum over repeated indices, the sum (\ref{eq sum FwFw}) can be written as
\begin{equation}
\label{eq no UA dependence in FwFw}
    U_{i'k'm',ik0}U^*_{i'\Tilde{k}'\Tilde{m}',ik_10}U^*_{j'k'm',jl0}U_{j'\Tilde{k}'\Tilde{m}',jl_10}=\sum_{\substack{i,j,i',j'}}F_{kl,ij}^U(i',j')F_{k_1l_1,i,j}^{*U}(i',j'),
\end{equation}
where the dependence on $U^A$ has vanished, and the expression depends only on the components of $U$. Therefore, using (\ref{eq no UA dependence in FwFw}) in equation (4), 
we have that $I_{B\rightarrow A}((U^A\otimes\mathbb{I}^{BC})\cdot U)=I_{B\rightarrow A}(U)$. Analogously, one finds the local invariance for $I_{A\rightarrow B}$.

\item $I_{B\rightarrow A}((\mathbb{I}^A\otimes U^B\otimes\mathbb{I}^{C}) U)=I_{B\rightarrow A}(U).$\\
According to Def 2, $I_{B\rightarrow A}=\int d\mu(V)\sum_{hfi'j'}\abs{\partial\rho'^A_{i'j'}/\partial \rho^B_{hf}}^2$. Alice's reduced final state is obtained tracing out systems $B$ and $C$ after the evolution, nevertheless both $(\mathbb{I}^A\otimes U^B\otimes\mathbb{I}^{C}) U$ and $U$ lead to the same  reduced state of Alice and therefore $\partial\rho'^A_{i'j'}/\partial \rho^B_{hf}$ remains unaltered, meaning that $I_{B\rightarrow A}$ is the same in both cases. 

\item $I_{B\rightarrow A}(U (U_0^A\otimes \mathbb{I}^{BC}))=I_{B\rightarrow A}(U).$

It follows from the (Haar) integration over all possible initial Alice's states. 

\end{enumerate}

\section{H. Proof of Property 3. $E[I_{A \rightarrow B}(U)]$}
From equation (4),
\begin{equation}
    I_{B \rightarrow A}(U)=\frac{1}{d_A(d_A+1)}D_{kl,hf}D_{k_1l_1,hf}(\delta_{i_1i}\delta_{jj_1}+\delta_{ij}\delta_{i_1j_1})F_{kl,ij}^U(i',j')F_{k_1l_1,i_1j_1}^{*U}(i',j'),
\end{equation}
and, applying the definition of the first moment,
\begin{equation}\label{eq EIU 12t}
\begin{split}
    &E[I_{B \rightarrow A}(U)]=\int d\mu(U) I_{B \rightarrow A}(U)\\&=\frac{1}{d_A(d_A+1)}\sum_{ i,j,i_1,j_1} D_{kl,hf}D_{k_1l_1,hf}(\delta_{i_1i}\delta_{jj_1}+\delta_{ij}\delta_{i_1j_1})\int d\mu(U)F_{kl,ij}^U(i',j')F_{k_1l_1,i_1j_1}^{*U}(i',j').
\end{split}
\end{equation}
Using the definition of the $F$ functions, the integral {over $U$} can be written as
\begin{equation}\label{eq EIU integrals}
    \sum_{k',\Tilde{k}',m',\Tilde{m}'}\int d\mu(U) U^*_{j'k'm',jl0}U^*_{i'\Tilde{k}'\Tilde{m}',i_1k_10}U_{i'k'm',ik0}U_{j'\Tilde{k}'\Tilde{m}',j_1l_10}=\sum_{k',\Tilde{k}',m',\Tilde{m}'}
    \braket{p_1p_2,q_1q_2}{r_1r_2,s_1s_2},
\end{equation}
where the integrals of the form $\int d\mu(V)V_{i_1j_1}^*...V_{i_pj_p}^* V_{k_1l_1}... V_{k_pl_p}$ have been denoted by\\ $\braket{(i_1...i_p),(j_1...j_p)}{(k_1...k_p),(l_1...l_p)}=:\braket{I,J}{K,L}$, and where, for short, it has been introduced the notation $p_1=j'k'm'$, $p_2=i'\Tilde{k}'\Tilde{m}'$, $q_1=jl0$, $q_2=i_1k_10$, $r_1=i'k'm'$, $r_2=j'\Tilde{k}'\Tilde{m}'$, $s_1=ik0$ and $s_2=j_1l_10$. The non-vanishing integrals are those such that $(r_1,r_2)=\sigma_1(p_1,p_2)$ and $(s_1,s_2)=\sigma_2(q_1,q_2)$, for $\sigma_1,\sigma_2\in S_2=\{id,\sigma\}$. Denote $P=(p_1,p_2)$, $Q=(q_1,q_2)$, $R=(r_1,r_2)$ and $S=(s_1,s_2)$, so that the order $2$ integrals are of the form $\braket{P,Q}{R,S}$ and we split them into the four (three non-equal) following (non-disjoint) sets of integrals: $\braket{P,Q}{P,S}$, $\braket{P,Q}{R,Q}$ and $\braket{P,Q}{P,Q}=\braket{PQ}{P_{\sigma}Q_{\sigma}}$. From set theory, given three sets $I_1$, $I_2$ and $I_3$ the number of elements of its union is 
\begin{equation}
    |I_1\cup I_2 \cup I_3|=|I_1|+|I_2|+|I_3|-|I_1\cap I_2|-|I_1\cap I_3|-|I_2\cap I_3|+|I_1\cap I_2\cap I_3|.
\end{equation}
Thus, we can write the order $2$ integrals in terms of the four (three non-equal) sets of integrals stated above:
\begin{equation}\label{eq integrals order 4 brakets}
\begin{split}
    \braket{P,Q}{R,S}&=
    \delta_{PR}\braket{P,Q}{P,S}+
    \delta_{QS}\braket{P,Q}{R,Q}+
    \delta_{PR_{\sigma}}\delta_{QS_{\sigma}}\braket{P,Q}{P_{\sigma},Q_{\sigma}}\\&
    -\delta_{PR}\delta_{QS}\braket{P,Q}{P,Q}
    -\delta_{PR}\delta_{PR_{\sigma}}\delta_{QS_{\sigma}}\braket{P,Q}{P,Q_{\sigma}}
    -\delta_{QS}\delta_{PR_{\sigma}}\delta_{QS_{\sigma}}\braket{P,Q}{P_{\sigma},Q}\\&
    +\delta_{PR}\delta_{PR_{\sigma}}\delta_{QS}\delta_{QS_{\sigma}}\braket{P,Q}{P,Q}.
\end{split}
\end{equation}
The integrals in (\ref{eq integrals order 4 brakets}), in terms of integrals depending on $p_1,p_2,q_1$ and $q_2$, are
\begin{enumerate}
    \item  $\braket{P,Q}{P,S}$, which can be written as 
    \begin{equation}\label{eq int PQ0}
    \begin{split}
        \braket{p_1p_2,q_1q_2}{p_1p_2,s_1s_2}&=\delta_{q_1s_1}\delta_{q_2s_2}\braket{p_1p_2,q_1q_2}{p_1p_2,q_1q_2}+\delta_{q_1s_2}\delta_{q_2s_1}\braket{p_1p_2,q_1q_2}{p_1p_2,q_2q_1}\\&
        -\delta_{q_1q_2}\delta_{q_1s_1}\delta_{q_1s_2}\braket{p_1p_2,q_1q_1}{p_1p_2,q_1q_1},
    \end{split}
           \end{equation}
    \item $\braket{P,Q}{R,Q}$, which takes the same value as $\braket{Q,P}{Q,R}$, and therefore it is computed as in 1,
    \item $\braket{P,Q}{P_{\sigma},Q_{\sigma}}=\braket{P,Q}{P,Q}=\braket{p_1p_2,q_1q_2}{p_1p_2,q_1q_2}$,
    \item $\braket{P,Q}{P,Q_{\sigma}}=\braket{p_1p_2,q_1q_2}{p_1p_2,q_2q_1}$, and
    \item $\braket{P,Q}{P_{\sigma},Q}$ which takes the same value as $\braket{Q,P}{Q,P_{\sigma}}$, and therefore it is computed as in 4.
\end{enumerate}
Therefore, we are left with the values of $\braket{p_1p_2,q_1q_2}{p_1p_2,q_1q_2}$ and $\braket{p_1p_2,q_1q_2}{p_1p_2,q_2q_1}$. From Fig. \ref{Figure order 2 integrals} we have that:
\begin{equation}\label{eq int PQ}
\braket{p_1p_2,q_1q_2}{p_1p_2,q_1q_2}=
    \begin{cases}
               \mathcal{I}_{(e)} & \text{if } p_1=p_2 \text{ and } q_1=q_2,\\
               \mathcal{I}_{(c)} & \text{if } p_1=p_2 \text{ and } q_1\neq q_2,\\
               \mathcal{I}_{(d)} & \text{if } p_1\neq p_2 \text{ and } q_1=q_2,\\
               \mathcal{I}_{(a)} & \text{if } p_1\neq p_2 \text{ and } q_1\neq q_2,\\
           \end{cases}
\end{equation}
and 
\begin{equation}\label{eq int PQ2}
\braket{p_1p_2,q_1q_2}{p_1p_2,q_2q_1}=
    \begin{cases}
               \mathcal{I}_{(e)} & \text{if } p_1=p_2 \text{ and } q_1=q_2,\\
               \mathcal{I}_{(c)} & \text{if } p_1=p_2 \text{ and } q_1\neq q_2,\\
               \mathcal{I}_{(d)} & \text{if } p_1\neq p_2 \text{ and } q_1=q_2,\\
               \mathcal{I}_{(b)} & \text{if } p_1\neq p_2 \text{ and } q_1\neq q_2.\\
           \end{cases}
\end{equation}
{Combining} (\ref{eq integrals order 4 brakets}), (\ref{eq int PQ0}), (\ref{eq int PQ}) and (\ref{eq int PQ2}), a generic integral of order $2$ can be written as
\begin{equation}\label{eq integrals order 4}
\begin{split}
\bra{p_1p_2,q_1q_2}&r_1r_2,s_1s_2\rangle=\delta_{p_1r_1}\delta_{p_2r_2}\Bigg( 
    \delta_{q_1s_1}\delta_{q_2s_2}\left\{\delta_{q_1q_2}\frac{1+\delta_{p_1p_2}}{D(D+1)}+(1-\delta_{q_1q_2})\left[\frac{\delta_{p_1p_2}}{D(D+1)}+\frac{1-\delta_{p_1p_2}}{D^2-1}\right]\right\} \\
&+\delta_{q_1s_2}\delta_{q_2s_1}\left\{\delta_{q_1q_2}\frac{1+\delta_{p_1p_2}}{D(D+1)}+(1-\delta_{q_1q_2})\left[\frac{\delta_{p_1p_2}}{D(D+1)}-\frac{1-\delta_{p_1p_2}}{D(D^2-1)}\right]\right\}\\
&-\delta_{q_1q_2}\delta_{q_1s_1}\delta_{q_1s_2}\frac{1+\delta_{p_1p_2}}{D(D+1)}\Bigg)\\
+&\delta_{q_1s_1}\delta_{q_2s_2}\Bigg(
    \delta_{p_1r_1}\delta_{p_2r_2}\left\{\delta_{p_1p_2}\frac{1+\delta_{q_1q_2}}{D(D+1)}+(1-\delta_{p_1p_2})\left[\frac{\delta_{q_1q_2}}{D(D+1)}+\frac{1-\delta_{q_1q_2}}{D^2-1}\right]\right\} \\
&+\delta_{p_1r_2}\delta_{p_2r_1}\left\{\delta_{p_1p_2}\frac{1+\delta_{q_1q_2}}{D(D+1)}+(1-\delta_{p_1p_2})\left[\frac{\delta_{q_1q_2}}{D(D+1)}-\frac{1-\delta_{q_1q_2}}{D(D^2-1)}\right]\right\}\\
&-\delta_{p_1p_2}\delta_{p_1r_1}\delta_{p_1r_2}\frac{1+\delta_{q_1q_2}}{D(D+1)}\Bigg)\\
+&\Big(\delta_{p_1r_2}\delta_{p_2r_1}\delta_{q_1s_2}\delta_{q_2s_1}-\delta_{p_1r_1}\delta_{p_2r_2}\delta_{q_1s_1}\delta_{q_2s_2}\Big)\Big(\delta_{q_1q_2}\frac{1+\delta_{p_1p_2}}{D(D+1)}+(1-\delta_{q_1q_2})\left[\frac{\delta_{p_1p_2}}{D(D+1)}+\frac{1-\delta_{p_1p_2}}{D^2-1}\right]\Big)\\
-&\delta_{p_1r_1}\delta_{p_1p_2}\delta_{p_1r_2}\delta_{q_1s_2}\delta_{q_2s_1}\left[\delta_{q_1q_2}\frac{2}{D(D+1)}+\frac{1-\delta_{q_1q_2}}{D(D+1)}\right]\\
-&\delta_{p_1r_2}\delta_{p_2r_1}\delta_{q_1s_1}\delta_{q_1q_2}\delta_{q_2s_2}\left[\delta_{p_1p_2}\frac{2}{D(D+1)}+\frac{1-\delta_{p_1p_2}}{D(D+1)}\right]\\
+&\delta_{p_1r_1}\delta_{p_1p_2}\delta_{p_1r_2}\delta_{q_1s_1}\delta_{q_1s_2}\delta_{q_1q_2}\frac{2}{D(D+1)},
\end{split}\end{equation}
and plugging it on equation (\ref{eq EIU integrals}) and subsequently in (\ref{eq EIU 12t}), we obtain the expected value of the measure of causal influence.  \\
Let $\emptybraket{...}$ denote the integral $\braket{p_1p_2,q_1q_2}{r_1r_2,s_1s_2}$ assuming that {the} indices are written in terms of the original ones, i.e.~$p_1=j'k'm'$, etc, and denote the same integral but with a certain index taking the same value {as} another index writing the equality of indices inside $\emptybraket{...}$ instead of the three dots, i.e.~the integral $\braket{p_1p_2,q_1q_2}{r_1r_2,s_1s_2}$ for $i_1=i$ will be written as $\emptybraket{i_1=i}$. Then, from equation (\ref{eq EIU 12t}), and denoting by $\lambda$ the set of indices $i'j'k'\Tilde{k'}m'\Tilde{m'}$,
\begin{equation}\label{eq EIU proof}
\begin{split}
     d_A&(d_A+1)E[I_{B\rightarrow A}(U)]=\sum_{\substack{\lambda iji_1j_1\\klk_1l_1hf}}D_{kl,hf}D_{k_1l_1,hf}(\delta_{i_1i}\delta_{jj_1}+\delta_{i,j}\delta_{i_1,j_1})\emptybraket{...} \\
     &=\sum_{\substack{\lambda ij\\klk_1l_1hf}}D_{kl,hf}D_{k_1l_1,hf}\emptybraket{i_1=i,j_1=j}+\sum_{\substack{\lambda ii_1\\klk_1l_1hf}}D_{kl,hf}D_{k_1l_1,hf}\emptybraket{j=i,j_1=i_1}.
\end{split}
\end{equation}
Since, for our purpose, the two terms in (\ref{eq EIU proof}) can be treated analogously, we will show the computations for the first term. We separate the cases where $h\neq f$ and $h=f$. In the former case {$D_{kl,hf}$} takes the value $\delta_{kh}\delta_{lf}$ and in the latter, $(-1)^{\delta_{k0}}(\delta_{h0}-1)\delta_{kl}\delta_{hf}$. Thus, the first term summed in the above expression can be written as
\begin{equation}\label{eq EIU proof tem1}
    \sum_{\substack{\lambda ij\\klk_1l_1h}}\sum_{f:f\neq h} \delta_{kh}\delta_{lf}\delta_{k_1h}\delta_{l_1f}\emptybraket{i_1=i,j_1=j}+\sum_{\substack{\lambda ij\\klk_1l_1hf}}\sum_{f:f=h}(-1)^{\delta_{k0}+\delta_{k_10}}(1-\delta_{h0})\delta_{hf}\delta_{kl}\delta_{k_1l_1}\emptybraket{i_1=i,j_1=j}.
\end{equation}

The first term of the sum (\ref{eq EIU proof tem1}) can be written as 
\begin{equation}\label{eq EIU term I}
\sum_{\substack{\lambda ijkl}}\emptybraket{i_1=i,j_1=j,k=k_1=h,l=l_1=f}-\sum_{\substack{\lambda ijk}}\emptybraket{i_1=i,j_1=j,k=k_1=l=l_1=f=h},
\end{equation}
and the second term is simplified as 
\begin{equation}\label{eq EIU term II}
    (d_B-1)\sum_{\substack{\lambda ij\\kk_1}}(-1)^{\delta_{k0}+\delta_{k_10}}\emptybraket{i_1=i,j_1=j,l=k,l_1=k_1}.
\end{equation}
Using \textit{Mathematica}, computing the integrals via (\ref{eq integrals order 4}), one obtains the sums in (\ref{eq EIU term I}) and (\ref{eq EIU term II}), e.g. the fisrt term in  (\ref{eq EIU term I}) is $\frac{d_A^2d_B(d_B^2d_C(d_A^2+d_C-1)-1)}{d_A^2d_B^2d_c^2-1}$, and (5)
is recovered. $E[I_{A \rightarrow B}(U)]$ is obtained by
permuting $A\leftrightarrow B$, i.e. $E[I_{A \rightarrow B}(U)](d_A,d_B,d_C)=\tau(E[I_{B \rightarrow A}(U)](d_A,d_B,d_C))$, where $\tau$ is the permutation $\tau(A)=B$, $\tau(B)=A$.

\section{I. Causal quantum switch}
Consider unitary evolutions $U^{A\substack{\nleftarrow\\[-1em]\rightarrow } B}$ and $U^{A\substack{\leftarrow\\[-1em]\nrightarrow } B}$ that permit a single influence direction indicated in their superscript. In analogy to the quantum switch one can create a unitary evolution $U_{sup}$ 
  that superposes the 
  CI $A\substack{\nleftarrow\\[-1em] \rightarrow} B$ and $A\substack{\leftarrow\\[-1em] \nrightarrow} B$, 
  controlled by an ancilla qubit $\ket{\chi^c}=\cos(\theta/2)\ket{0}+e^{i\phi}\sin(\theta/2)\ket{1}$, for $\theta\in[0,2\pi)$ and $\phi\in[0,\pi)$,  
\begin{equation}
U_{sup}=\ketbra{0}{0}\otimes U^{A\substack{\nleftarrow\\[-1em]\rightarrow } B}+\ketbra{1}{1}\otimes U^{A\substack{\leftarrow\\[-1em]\nrightarrow } B}.\label{eq.9}
\end{equation}
  Fig \ref{Plot influence superposition} 
  shows the 
  CI from $A$ to $B$ and vice-versa, with $d_A=d_B=2$,  for $U_{sup}^1=\ketbra{0}{0}\otimes U^{(123)}+\ketbra{1}{1} \otimes U^{(132)}$ depending on the control qubit, where one sees that if $\ket{\chi^c}=\ket{0}$, $I(U_{sup},\ket{\chi^c})=I(U^{(123)})$ and if $\ket{\chi^c}=\ket{1}$, $I(U_{sup},\ket{\chi^c})=I(U^{(132)})$, for $I\in\{I_{A\rightarrow B},I_{B\rightarrow A}\}$. 
\begin{figure}[h]
    \centering
    \includegraphics[width=77mm]{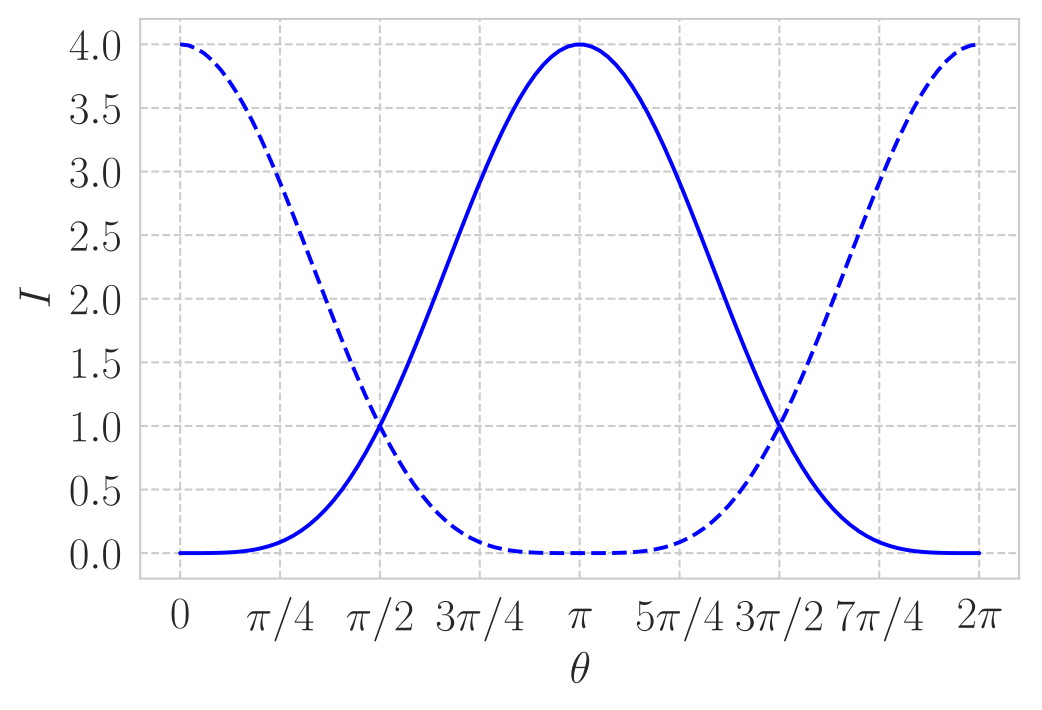}
    \caption{ (Color online) $I_{A \rightarrow B}$ (dashed line) and $I_{B\rightarrow A}$ (continous line) of $U_{sup}^1$ depending on the control qubit $\ket{\chi^c}=\cos(\theta/2)\ket{0}+\sin(\theta/2)\ket{1}$, parametrized by $\theta$. 
    }
    \label{Plot influence superposition}
\end{figure}
Classically, CI is usually considered one-way only (represented by an arrow in a directed acyclic graph), as time-ordering of events and forward-in-time-only causal influence is assumed. In \eqref{eq.9}, both CI are forward in time.  When time stamps of events can be exchanged, mixtures of causal influences     
are also possible classically (lung cancer can be caused by smoking, but also incite people to enjoy some final cigarettes...), but the superpositions introduced here go beyond this.   More generally, the four possible CI  
options could be superposed, and, for open quantum systems, Kraus-operators with different causal influence.

\section{J. Propagation of causal influence}
{For the physical example
  of two double quantum dots (DQD) at distance $l$ from each other coupled with dipole interaction to
  black-body radiation, with a UV frequency cut-off $y_m=\omega_{max}\tau $, where $\tau=\hbar/k_BT$ and $T$ is the temperature,
the functions $f(t)$ and $\varphi(t)$ can be {obtained} analytically if one approximates coth $\simeq1$ for $y_m\gg1$ [55].
} 
The
argument of the $\sin$ in (6) 
becomes 
\begin{equation}
\begin{split}
    \frac{At}{t_0^3}\{-2\sin(y_{m}t_0)+\Si[y_{m}(t-t_0)]+2\Si(y_{m}t_0)\\-\Si[y_{m}(t+t_0)]\}+\frac{2A}{y_{m}t_0^3}\sin(y_{m}t)\sin(y_{m}t_0),    
\end{split}\label{eq.12}
\end{equation}
where $t_0=l/c$ denotes the time of travel of a light signal between
the two DQD and  $A=\alpha_0d^2/\pi c^2\tau^2$, $d$ is the
dipole moment of the quantum system divided by the electron
charge, and
$\alpha_0\simeq\frac{1}{137}$ and $c$ are the fine-structure constant
and the speed of light in vacuum, respectively. In \eqref{eq.12}  and onwards, both $t$ and $t_0$ are in units of $\tau$. 
The exponent in (6) 
becomes
$-\alpha_0d^2\omega_{max}^2/3c^2$. 
\end{widetext}

\end{document}